\documentclass[twocolumn,prl]{revtex4-1}
\usepackage{graphicx}
\usepackage{amsmath}
\usepackage{amssymb}
\usepackage{bm}

\begin{document}

\noindent {\bf Comment on \textquotedblleft Space-Time Crystals of Trapped Ions\textquotedblright: And Yet it Moves \emph{Not}!}

\vspace*{0.5\baselineskip}







In a recent Letter \cite{TongcangLi2012}, Tongcang Li \emph{et al.} (TL) proposed an experimental realization of Wilczek's concept of \textquotedblleft quantum time crystals" \cite{Wilczek2012}, defined as systems which, in their quantum mechanical ground state, exhibit a periodic oscillation of some physical observable. As discussed in Ref.~\cite{Bruno2012}, Wilczek's proposal is actually erroneous, and his model does \emph{not} constitute a \textquotedblleft quantum time crystal". In the present Comment, I point out that TL's proposal is incorrect too.

The model proposed by TL is a variant of Wilczek's model, and consists of an assembly of $N$ ions of mass $M$ confined in a one-dimensional ring-shaped trap threaded by an Aharonov-Bohm (AB) magnetic flux. TL claim that the Wigner crystal resulting from Coulomb repulsion among the ions will, in its ground state, exhibit a periodic, time-dependent behavior, consisting in a rotational motion, due to the AB flux . They argue that this zero-point rotational motion can be observed by \textquotedblleft tagging" one ion by promoting it to an excited hyperfine level, and subsequently monitoring the periodic passage of the tagged ion.

The claims made by TL are problematic is various ways. In the first place, one observes that if the Wigner crystal, in its ground state, would be really breaking time translational symmetry, as TL claim, it would generate a time-dependent electromagnetic field and radiate energy, thereby violating the law of energy conservation. The authors' statement that \textquotedblleft since the ions are in the ground state already, there is no radiation loss due to the rotation", is tautological, and does not explains how rotating localized charges could possibly bypass Maxwell's equations and avoid radiating.

On the other hand, it is not difficult to see that the ground state of TL's Hamiltonian (Eq.~[1] of \cite{TongcangLi2012}) has charge and current densities which do not break the space rotational symmetry and is perfectly time-independent. Thus, although the (finite size) unpinned Wigner crystal freely rotates, thereby generating a measurable current, it is actually a \textquotedblleft floating crystal'' \cite{Mikhailov2002,Drummond2004} with stationary and rotationally uniform density, and does not constitute a \textquotedblleft quantum time crystal" as defined above. The attempt of TL to elude this problem, namely, the proposition of hyperfine-tagging one ion, does not help: the ground state of the new system, comprising $N\! \! -\!\! 1$ untagged ions plus one tagged ion, still has a rotationally uniform and time-independent ground state. That the ground state expectation value $\langle \mathcal{A}\rangle_0$ of any observable $\mathcal{A}$ is time-independent, for a \emph{finite} system, follows trivially from Ehrenfest's theorem and from the fact that the ground state is an eigenstate of the Hamiltonian: $\frac{\mathrm{d}}{\mathrm{d}t}\langle \mathcal{A}\rangle_0= \frac{\mathrm{i}}{\hbar}\langle\Psi_0 | [H,\mathcal{A}] | \Psi_0\rangle =0$; this is a hard and well known fact that cannot be circumvented.

The above failure can be traced back to the fact that what TL are looking for is a spontaneous breaking of symmetry, which, as is well known, can take place only in the thermodynamic limit (i.e., $N\to\infty$) and cannot be captured by a calculation for finite $N$ as done by TL. More precisely, in order to describe a spontaneous symmetry breaking, one needs to introduce a small symmetry-breaking external potential $v$, and take the limit $N\to\infty$ \emph{before} taking the limit $v\to 0$ \cite{Bogoliubov1970}. TL effectively take limits in the wrong order and thus obtain a state with unbroken symmetries. Thus the calculations presented by TL cannot usefully describe whether or not the Wigner crystal with broken symmetry rotates under the effect of the AB flux.

If one introduces a small symmetry-breaking (pinning) potential $v$, the motion of the ions around the ring has to take place via tunneling. Because the one-dimensional Wigner crystal (of lattice parameter $a$) is ideally stiff (i.e., the sound velocity is infinite \cite{Schultz1993}), the tunneling involves the coherent motion of all $N$ ions, and the effective tunneling mass is $N M$; as a result the tunneling probability, and hence the resulting current, contains the factor
$A(v,N) \simeq \exp \left({-\frac{a}{\hbar}\sqrt{2 N M v}}\right)$ \cite{Krive1995}. The correct thermodynamic limit yields $\lim\limits_{v \to 0} \lim\limits_{N \to \infty} A(v,N) =0$, implying that the Wigner crystal is completely insensitive to the AB flux, in sharp contrast with TL's improper limit: $\lim\limits_{N \to \infty} \lim\limits_{v \to 0} A(v,N) =1$. Physically, the insensitivity to the AB flux (i.e., to the twisted boundary condition for the phase of the wavefunction) of the state with spontaneously broken symmetry is due to the localization of the wavefunction (the value of the phase is immaterial as the amplitude vanishes), as discussed in \cite{Bruno2012}, and first pointed out long ago by W.~Kohn in his classic paper \textquotedblleft Theory of the Insulating State" \cite{Kohn1964}.

I am grateful to Philippe Nozi{\`e}res for helpful discussions.

\vspace*{0.5\baselineskip}

\noindent Patrick Bruno\\
{\small European Synchrotron Radiation Facility, BP 220, F-38043 Grenoble Cedex, France} \\
{ } \\

\end{document}